# Towards Semantic Web: Challenges and Needs

*Bryar Hassan[1], Srinandan Dasmahapatra[2]*

[1]Kurdistan Institution for Strategic Studies and Scientific Research, Directorate of Information Technology,
Shorsh Str., Sulaimani, Kurdistan Region, Iraq
*bryar.hassan@kissr.edu.krd*

[2]School of Electronics and Computer Science, Southampton University,
Southampton, SO17 1BJ, UK
*sd@ecs.soton.ac.uk*

**Abstract:** There has been recently a growth of interest in developing the current machine-readable Web towards the next generation of machine-understandable Web - Semantic Web. The development of the Web to a global business was reasonably fast, whereas Semantic Web development has taken time from a plan to be used as the mainstream Web. It is also important to note that the use of Semantic Web would only be successful in small technologies. However, the goal of Semantic Web is to be used in big technologies and to be the mainstream Web. Some challenges may impede make further progress of Semantic Web. In this review paper, an overview of the current status and future needs of Semantic Web will be presented. Specifically, the challenges and needs of Semantic Web in the hope of shedding some light on the adoption or infusion of Semantic Web in the future will be discussed. Then, a critical evaluation of these challenges and needs will be presented. Semantic Web has a clear vision. It is moving, in line with this vision, towards overcoming the challenges and usability in real world applications.

## 1. Introduction

When The World Wide Web (WWW) revolutionised the sphere of information technology. However, most of the current unstructured and semi-structured contents of the Web are machine-readable rather than ma-chine-understandable. Semantic Web was introduced with its vision by Tim Berners-Lee in 2001 as the complement of Web 3.0 in order to improve the current Web from machine readable into machine understandable [1]. The goal of Semantic Web is to develop languages so as to de-scribe and structure information on the Web to be understandable by ma-chine. There have been many researches and attempts on Semantic Web development to realise its goal, but Semantic Web has not come true yet. In addition, this development has taken time in order for Semantic Web to be the mainstream Web, compared to the development of the Web, which was reasonably fast. The question raised why has it not been the mainstream Web yet? There are some challenges facing Semantic Web that impede its progression. This paper will present the vital challenges of Semantic Web with their possible solutions for the future to move for-ward from the today's Web into tomorrow's Web of Semantics.

### 1.1 Semantic Web Brief History

In 1989, the World Wide Web was proposed as a development project to CERN by Tim Berners-Lee [1]. After two years, a portable browser was distributed. Then, a commercial browser flourished in 1991. By the end of 1995, Internet Explorer was released and W3C had been established as a standard body for the Web. Due to encouragement to add semantic meaning to web pages, the idea of Semantic Web was initiated in 1996 to automate everything on the Web [1]. Initially, the first draft of Resource Description Framework (RDF) for defining metadata was available in 1997. Then, a roadmap was published for Semantic Web as a notion be-yond metadata in 1998 that includes query languages, inference rules and proof validation by Tim Berners-Lee [1]. Next, the use of metadata (de-fined as a data about data) was proposed as a solution and became a W3C recommendation in 1999. Furthermore, the vision of Semantic Web including trust was extended in 2003 by Tim Berners-Lee [2].

### 1.2 Semantic Web Architecture

Semantic Web architecture was proposed by Tim Berners-Lee in 2001 and it has been standardised as a component of the Web 3.0 by W3C [3]. URI and Unicode as foundation layer allow objects to be addressed by unique identifiers and ensure that technologies are applicable to all languages. XML and XML Schema are used as information presenter, whereas RDF as a language for data model representation is used to make metadata and comments. The upper layer of RDF is the Web Ontology Language (OWL), which is a family of knowledge representation languages. It can add more vocabulary than RDF for property and class descriptions. Above the ontology layer, there are trust and proof layers for inference support. Semantic Web layers are related together and they are accomplishments of each other.

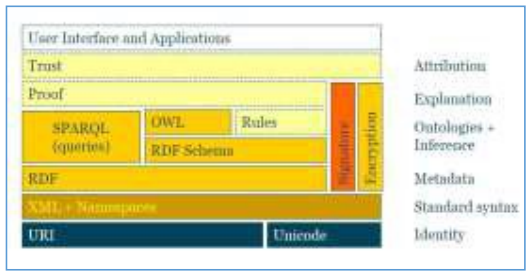

Figure 1: Semantic Web Architecture (adapted from [4])

### 1.3 Semantic Web Current Status

One of the important bases of Semantic Web is its architectural layered model representing a model for current and emerging technologies of Semantic Web. More importantly, Semantic Web is a standardisation of W3C. Ontologies have given more stability and improvement by standardising RDF and OWL [3]. This standardisation gives more opportunity to Semantic Web to move forward, fit into industry standards, and build next generation applications. Most importantly, using Semantic Web in small technologies is successful. In this respect, Semantic Web looks re-assuring to come true entirely, yet falsely so. The main aim of Semantic Web is to be used in big technologies and to be the mainstream Web, but it has not used for either of them yet. There are many challenges that should be pointed out before Semantic Web can be envisioned completely.

## 2. Challenges and Needs

There are many challenges confronting Semantic Web. This section will outline six main serious challenges with their possible needs and solutions. These challenges should be addressed before the complete vision of Semantic Web, if ever, is to be envisaged.

### 2.1 Ontologies

The most important challenge of Semantic Web is ontology language and its related aspects. At first blush, ontological models are constructed by OWL. Each model is like an abstraction which creates for an objective. Accordingly, there are issues such as assumptions for building the modelling process, methodologies for moving requirements to models, and limitations of the models [2]. These issues should be addressed be-fore the reach of ontological modelling to entity-relationship, and object or process modelling. Second, designing and expressing ontologies are not always perfect [5, 6]. Designing ontologies which is known as knowledge representation language of Semantic Web is not always fine enough and it can be assessed based upon real applications. Based on that, there are some issues that should be addressed such as ontology integration, ontology mapping, ontology translation, ontology consistency check, and ontology reuse. The future requirements to deal with this issue should include mapping reuse and ontology integration, standard development ontology for different fields, and ontology integration with applying time notion. Similarly, there is still the rigour of expressing ontologies. Particularly, this expression for complex languages such as OWL is more difficult to handle for both machines and humans than for light-weight languages such as RDF [6]. Third, ontology aging is another issue that should be dealt with in order to enable evolvement of ontologies with environmental changes [5]. In other words, ontologies should have flexibility to adapt themselves with environmental parameters such as time. Finally, it has been suggested that a vital effort should be made to create common widely used ontology for Semantic Web. Likewise, the kernel ontology should be used by all domains [7].

### 2.2 Multilingualism and Ontology Translation

Multilingualism is one of the most important challenges of Semantic Web [5, 7]. Although the English language is considered as the predominant language for Web contents, there are still vital documents on the Web which are written in other languages. This plays an important role at both the ontology level, and annotations and user interface levels. Necessities for the future with regard to internationalisation are development of a native ontology tool, ontology mapping and translator in different languages, and culture and national ontology integration. Some other issues of Semantic Web are cultural requirements and integration with languages [5]. People may think in different ways about a single concept based upon their historical and cultural background. Thus, tools should be created by different languages or developing ontology translators that map a content to other languages. More importantly, ontologies have problem with translating people's emotional expressions in all languages [2]. There is not only one way of speaking the English language. For instance, every human speaks in his or her own way. People have different ways of speaking and communicating with one another. Only syntax and semantics can be defined by programming languages. This individuality may cause restrictions for users to express their intentions. An approach of translation between ontologies has been recently taken so as to transform ontologies to common syntax address. This is a sensible start, yet metaphoric reasoning may be needed due to huge semantic differences.

### 2.3 Trust and Proof

Semantic Web may face another challenge when it is used for killer applications due to trust and poof layers [5]. Trust and proof layers have not been taken into account as the existing applications of Semantic Web are, in general, context dependent. Nevertheless, there will be different applications of Semantic Web in the future and this will presumably pose vulnerability for Semantic Web systems. Among the other challenges, trust and proof checking mechanisms, and digital signatures should be ad-dressed for future work. More significantly, data credibility, metadata control, and data mash-up privacy implication remain open in Semantic Web [8]. As a penetration of Web 2.0, these issues are still existent in Se-mantic Web as a component of Web 3.0.

### 2.4 Scalability

Managing its content in a scalable manner is another challenge of Se-mantic Web [6, 7]. This challenge is related to two underlying issues: first is storing Semantic Web content, and the second is a mechanism to find information with ease. To cope with these challenges, essential effort should be conducted to store Semantic Web contents and provide a mechanism to find the contents in a scalable manner. Solutions should be provided when Semantic web emerges completely. In accordance with [8], the scalability issue of Web 2.0 will penetrate in Web 3.0's applications. Scalability is still existent in the current Web 2.0. Accordingly, scalability in Semantic Web as a component of Web 3.0 will pose a challenge for professionals creating applications.

### 2.5 Security

Security is considered as another challenge for Semantic Web [6, 8, 9]. There are some security issues in Web 2.0 that have not been solved yet. These issues are possibly penetrated to Web 3.0's applications because Web 3.0 is considered as the extension of Web 2.0 and hence Semantic Web as a component of Web 3.0 may face the same issues. In this con-text, security challenges such as data credibility, metadata control, data mash-up privacy implication remain open in web 3.0. More significantly, Semantic Web and its model layers are not fully secured [9]. All Semantic Web components such as XML, RDF, and information should be se-cured in order for Semantic Web to be secured accordingly.

### 2.6 Application and Usability

Semantic Web applications and related aspects are issues that researchers always focus on [5]. Small applications of Semantic Web have been already developed, whereas killer ones are under development. Some challenging applications should be fulfilled that may help the way of using our computers such as "Augmented Personal Memories", world knowledge integration, and Semantic Web services integration with common used application. In addition, it has been argued that the essential challenges facing Semantic Web's applications are usability and visualisation [6, 7]. The ease of use for both developer and user interfaces are essential. In order to be able to use current Semantic Web tools, Semantic Web and Web standard expertises may be needed. It is viable that Semantic Web tools and applications should be simple and easy to use by both users and developers. Meanwhile, to accommodate huge amount of information in less space in Semantic Web applications, a new technique of hypertext visualisation for visualising Semantic Web may be required. In turn, users can easily interact with Semantic Web applications in a friendly way.

## 3. Evaluation

Six major challenges with some possible solutions for the future of Semantic Web have been presented in this paper. One question needs to be asked, however, as to peculiarity and precision of these challenges and their solutions. Another question would be on the extent to which these challenges are critical.

### 3.1 Ontologies

There is no doubt that ontologies and related aspects are the key challenges of Semantic Web. Nonetheless, researchers may have different perspectives on this challenge.[2] mentions some issues to be addressed such as assumptions for building the modelling process, methodologies for moving requirements to models, and limitations of the models. Similarly, all of these challenges are addressed in [10] to be solved in the future. In regard with designing and expressing ontologies[5, 6], ontologies are not always perfect enough. This issue relates to mostly writing and designing ontologies rather than ontologies themselves. Based on that, the way of designing and writing ontologies may need to be improved. Ex-pressing ontologies either for complex language or light-weight language, on the other hand, is always difficult and challenging for researchers. More recent arguments agreed with [7] on ontology usage that ontologies should be developed for different domains and applications [11]. Like-wise, companies and governments should be motivated to release ontology designers to share and build domain descriptions [12].

### 3.2 Multilingualism

Multilingualism, ontology translation, and culture requirement integration are the issues of Semantic Web. These issues could be dealt with and solved in the future without any difficulty. Nevertheless, the suggested solution is not precise enough because it has not mentioned how exactly it should be dealt with.

### 3.3 Trust and Proof

It has been contended that the future applications of Semantic Web will not probably be compatible with current trust and proof layers of Semantic Web [5]. This may be considered as a future possibility rather than an issue of Semantic Web as what do the future applications of Se-mantic Web look like is not yet known. They may not be different from the current applications of Semantic Web.

### 3.4 Scalability and Security

Referring to [8], scalability and security issues of web 2.0 will penetrate in web 3.0's applications. This is a relatively strong assertion be-cause of lack of supporting evidence. Web 3.0 is an advanced version of Web 2.0 to solve the current problems of Web 2.0. Therefore, Web 3.0 may have significantly big advantage over Web 2.0. [13] concludes that Web 3.0 can bring new spectacular applications in comparison to Web 2.0 with the similar magnitude of Web 1.0 and Web 2.0 separation. Thus, to a lesser extent, Web 2.0 may have negative effect on Web 3.0 and Semantic Web, yet it has positive effect on Semantic Web to a more ex-tent.

### 3.5 Application and Usability

Subject to application and usability issues [5, 6], it has been argued that applications of Semantic Web should be free and used widely by users. Difficulties arise, however, when an attempt is made to use Semantic Web applications since there is dearth of applications [6]. Additionally, Semantic Web is used currently for small applications rather than killer ones. Small applications may not really represent Semantic Web since they are not compliant with all features of Semantic Web and it has not been known yet that small and killer applications will be either similar or different from one another. Significantly, all applications of Semantic Web should be freely accessible online and described to users [14]. An-other critique of these papers is that they have not mentioned how to build the killer applications and how they should be made. [12] points out that companies and governments should be motivated to build Se-mantic Web-based applications and applications of Web 3.0 should be extended to browsers or other Web tools.

Thus, some of these challenges are critical, while some others are small issues. Similarly, the solutions are partly accurate,

whereas the other part is not significantly accurate and it is merely an expectation for the future needs of Semantic Web. These challenges may all hinge on one another and solving all of them may sometimes have a daunting proposition.

## 4. Conclusion and Future Direction

This paper has provided an overview of Semantic Web and a research review of its challenges and requirements for the future. First, it has introduced Semantic Web by briefly reviewing history, architecture and current status. Next, it has presented six of the most crucial challenges and solutions for Semantic Web. Final, it has also evaluated Semantic Web challenges and solutions. Furthermore, Semantic Web is inevitable and it is not lost cause as it has a clear vision and is currently in use for small applications. Nonetheless, all of the issues of Semantic Web should be resolved based on the suggested accurate solutions in this paper so as to scale beyond its own stronghold. Incidentally, development of many technologies, applications, policy and legislative frameworks may be needed to envision Semantic Web. Sooner or later, Semantic Web will be envisaged. Regardless of the extent of its success, its challenges should be addressed to make it ready for use in killer applications in move into the mainstream Web.

## References

[1] T. Berners-Lee, "Information management: A proposal," 1989.
[2] M. Wilson and B. Matthews, "The semantic Web: prospects and challenges," in *Databases and Information Systems, 2006 7th International Baltic Conference on*, 2006, pp. 26-29.
[3] T. Berners-Lee, J. Hendler, and O. Lassila, "The semantic web," *Scientific american,* vol. 284, pp. 28-37, 2001.
[4] I. Horrocks, B. Parsia, P. Patel-Schneider, and J. Hendler, "Semantic web architecture: Stack or two towers?," in *Principles and practice of semantic web reasoning*, ed: Springer, 2005, pp. 37-41.
[5] A. Andjomshoaa, F. Shayeganfar, and R. Wagner, "Semantic Web challenges and new requirements," in *Database and Expert Systems Applications, 2005. Proceedings. Sixteenth International Workshop on*, 2005, pp. 1160-1163.
[6] A. Ankolekar, M. Krötzsch, T. Tran, and D. Vrandecic, "The two cultures: Mashing up Web 2.0 and the Semantic Web," in *Proceedings of the 16th international conference on World Wide Web*, 2007, pp. 825-834.
[7] R. Benjamins, J. Contreras, O. Corcho, and A. Gomez-Perez, "The six challenges of the Semantic Web," 2002.
[8] M. M. I. Pattal, Y. Li, and J. Zeng, "Web 3.0: A real personal web! More opportunities and more threats," in *Next Generation Mobile Applications, Services and Technologies, 2009. NGMAST'09. Third International Conference on*, 2009, pp. 125-128.
[9] B. Thuraisingham, "Security issues for the semantic web," in *null*, 2003, p. 632.
[10] N. Shadbolt, W. Hall, and T. Berners-Lee, "The semantic web revisited," *Intelligent Systems, IEEE,* vol. 21, pp. 96-101, 2006.
[11] V. R. Benjamins, "Near-term prospects for semantic technologies," *Intelligent Systems, IEEE,* vol. 23, pp. 76-88, 2008.
[12] J. Hendler, "Web 3.0: Chicken farms on the semantic web," *Computer,* vol. 41, pp. 106-108, 2008.
[13] J. Cardoso, "The semantic web vision: Where are we?," *Intelligent Systems, IEEE,* vol. 22, pp. 84-88, 2007.
[14] M. Klein and U. Visser, "Guest Editors' Introduction: Semantic Web Challenge 2003," *IEEE Intelligent Systems,* pp. 31-33, 2004.

## Author Profile

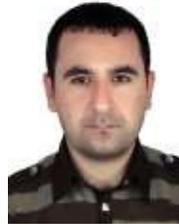

**Bryar Hassan** received BSc Engineering degree in Software Engineering from Salahaddin University-Erbil in 2007 and MSc Software Engineering in 2013 from the University of Southampton, UK. He is currently a lecturer/researcher in Kurdistan Institution for Strategic Studies and Scientific Research as we as the director of Information Technology Directorate in Kurdistan Institution. His research interests are in Semantic Web, Web 3.0, Information Systems, and E-Government.

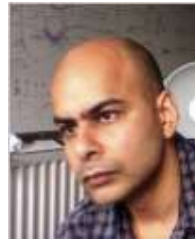

**Dr Srinandan Dasmahapatra** has received PhD in Computer Science from the University of Southampton, UK. He is currently an academic staff in in Vision, Learning and Control from the University of Southampton. His research interests are in Computational Biology, Machine Learning, Artificial Intelligence, and Intelligent Systems.